\documentclass{sig-alternate-05-2015}

\usepackage{filecontents,lipsum}
\usepackage[noadjust]{cite}
\usepackage{times}
\usepackage{helvet}
\usepackage{courier}
\usepackage[utf8]{inputenc}

\usepackage{algorithm}
\usepackage[noend]{algpseudocode}

\usepackage{spreadtab}

\usepackage{amsmath,amssymb,amsthm,amsfonts}
\usepackage{graphicx}
\graphicspath{{figs/}}
\usepackage{xspace}

\usepackage{subfig}

\newcommand{\myalgo}{\mbox{\sc PTR}\xspace}
\newcommand{\myalgotime}{\mbox{\sc TPTR}\xspace}

\newcommand{\tweets}{\ensuremath{{\cal T}}\xspace}
\newcommand{\users}{\ensuremath{{\cal U}}\xspace}
\newcommand{\topics}{\ensuremath{{\cal C}}\xspace}
\newcommand{\htags}{\ensuremath{{\cal H}}\xspace}
\newcommand{\cusers}{\ensuremath{{\cal Z}}\xspace}

\newcommand{\pd}{IT13\xspace}
\newcommand{\eu}{EU14\xspace}

\newcommand{\datapd}{IT13\xspace}
\newcommand{\dataeu}{EU14\xspace}

\begin{document}
\CopyrightYear{2016} 
\setcopyright{acmlicensed}
\conferenceinfo{SIGIR '16,}{July 17 - 21, 2016, Pisa, Italy}
\isbn{978-1-4503-4069-4/16/07}\acmPrice{\$15.00}
\doi{http://dx.doi.org/10.1145/2911451.2914716}


\clubpenalty=10000 
\widowpenalty = 10000

%

\title{Polarized User and Topic Tracking in Twitter}

%
\numberofauthors{4} 
%
\author{
%
%
\alignauthor
Mauro Coletto\\
       \affaddr{IMT Lucca, ISTI--CNR Pisa}\\
       \email{mauro.coletto@isti.cnr.it}
\alignauthor
Claudio Lucchese\\
       \affaddr{ISTI--CNR Pisa}\\
       \email{claudio.lucchese@isti.cnr.it}
\and
\alignauthor 
Salvatore Orlando\\
       \affaddr{DAIS--University of Venice}\\
       \email{orlando@unive.it}
\alignauthor Raffaele Perego\\
       \affaddr{ISTI--CNR Pisa}\\
       \email{raffaele.perego@isti.cnr.it}
}

\maketitle

\begin{abstract}
Digital traces of conversations in micro-blogging platforms and OSNs provide information about user opinion with a high degree of
resolution. These information sources can be exploited to understand and monitor 
collective behaviors. 
In this work, we focus on {\em polarization classes}, i.e., those topics that
require the user to side exclusively with one position. The proposed method
provides an iterative classification of users and keywords: first, \textit{polarized users}
are identified, then \textit{polarized keywords} are discovered by monitoring the
activities of previously classified users. This method thus allows tracking users and topics over time.
We report several experiments conducted on two Twitter datasets during
political election time-frames. We measure the user classification accuracy on a golden set of users, and analyze the relevance of the extracted keywords for the ongoing political
discussion.
  
%
%
%
%
%
%
%
%

\end{abstract}




\section{Introduction}
\label{sec:intro}

Recently, the analysis of blogging platforms and streaming
information sources (e.g., Twitter) has received great attention in the Information Retrieval
and in the Data Mining communities. 
We focus on the frequent scenario where users interact and produce contents
according to a set of {\em polarization classes}. By polarization classes we mean
subjects that require the user to side exclusively with one part. Political
parties are typical examples of these classes. 
Other examples include brand analysis, products
comparison, and opinion mining in general. In these scenarios the polarization classes are known, and some limited
information may also be available, e.g., a set of relevant keywords. This
limited knowledge allows us to restrict the scope of the analysis, but several
challenging tasks are left open. The first is how to identify the users being
polarized (or not) according to those classes. The
second task concerns the identification of the most relevant sub-topics being discussed among such users. The third is how to monitor the evolution of such user
communities and their on-line discussions over time. Those tasks are all very
challenging as the available knowledge may be approximate or insufficient, and
it may also become obsolete over time. Therefore, the classification into
polarization classes should be able to self-update continuously by catching
upcoming relevant users and discussion topics. 
The present work is related to the
Topic Detection and Tracking (TDT) subject \cite{allan2012topic}, which has been
widely explored within the scope of news stream analysis \cite{walls1999topic}. 
We focus on content and user tracking for polarized users. This notion is connected with the concept of controversy in Social Media, which have been studied, mostly in political contexts, using data coming from different sources \cite{adamic2005political, garimella2015quantifying,makazhanov2014predicting,van2015determining}.
Another related research area is trending topics analysis.
Various trend detection models are proposed in \cite{mathioudakis2010twittermonitor, lin2011smoothing}.
Our approach is different in several regards from current literature, since we rather focus on the
identification of polarized communities.
In our experiments we use electoral data from Twitter. In this case, the polarization classes are political parties and
candidates. 
Several works  analyzed the opportunities
and limitations in using Twitter as a predictor of an election's outcome
\cite{tumasjan2010predicting, coletto, gayo2011limits}.
Our goal is completely different, as we do not draw any conclusion about the
expected share of votes for the given parties or candidates. We use this
specific kind of data, as it is a typical example of polarized users. We
show that the proposed algorithm is able to identify polarized users, by also
analyzing the ongoing discussions among the respective communities.
The main contribution of this work is a new iterative algorithm, named
\myalgo (Polarization TRacker), for the discovery of {\em polarized users} in a Twitter stream, and a temporal version \myalgotime (Temporal \myalgo), 
able to track users and topics over time. While
there exist several works about community detection and trending topic
tracking, we propose a novel setting where the number of communities is
known, but very little information is provided (a keyword per class only),
and those communities are competing with each other. 
We conduct an objective
evaluation of the proposed algorithms by measuring their classification accuracy on a {\em
golden} set of users. 



\section{User and Topic Tracking} \label{sec:problem}

Let $\tweets = \{t_1,
t_2, \ldots\}$ be the stream of tweets generated by the set of users
$\users=\{u_1, u_2, \ldots \}$. 
We focus on the analysis of user behavior 
with respect to a set of {\em polarization classes} \topics. 
The goal of this work is thus to build a \emph{partitional clustering} of the Twitter users, where each of the clusters is associated by construction with a single polarization class (or unassigned). 
Our method can be seen as a \emph{semi-supervised clustering} one, although, unlike classic methods, we do not provide any class representative around which the final clustering is induced. Indeed, the proposed method
is only {\em loosely} supervised as the only knowledge available is the number of classes,
and a short class description (a keyword).


An important issue is the evaluation of our algorithm. To this end, we exploit a 
{\em golden set} of polarized users, each unequivocally associated with a class $c \in \topics$. 
Note that such knowledge is not exploited to train a classifier, but only
for evaluation purpose.



\subsection{The \myalgo algorithm}

The Polarization TRacker (\myalgo) algorithm requires some initial {\em seed topics} that identify the classes of interests.
We propose to identify them with 
a single textual keyword for each class $c \in \topics$. Although each keyword identifies a topic, e.g., a political party,  it is not sufficient to correctly classify users, as
all these seed topics are likely to be mentioned in many users' tweets, e.g., to contrast the achievements of a given party with the deficiencies of the others.
Without loss of generality, we limit our keyword 
selection to Twitter {\em hashtags}. 
Therefore, the single textual keyword we initially choose for each class $c$ is a single hashtag appearing in the user tweets, and around them we start identifying the user clusters. The final goal is to extract the best discriminating hashtags that are able to identify the actual clusters of polarized users, who belong with high probability to one of the classes $c \in \topics$. 

We denote the representative hashtags, one for each $c \in \topics$, called \emph{seed hashtags}, by $H_c^{\tau=0}$, where
$\tau$ is the algorithm's iteration number.  
Note that each initial set $H_c^{\tau=0}$, one for each $c$, is not necessarily composed of a discriminating hashtag. 
This set $H_c^{\tau=0}$ is then used to classify polarized users on the basis of their
use of the seed hashtags. We denote by $U_c^{\tau+1}$ the clusters of users in
\users that are identified as belonging to class $c$, according to their tweets
and to the given hashtags $H_c^{\tau}$. Similarly, the new hashtags $H_c^{\tau+1}$
are generated by finding those that best discriminate the users
in $U_c^{\tau+1}$. This refinement process is iterated for all $c \in \topics$: from hashtags $\{H_c^\tau\}_{c \in \topics}$ to users $\{U_c^{\tau+1}\}_{c \in \topics}$, and finally to hashtags $\{H_c^{\tau+1}\}_{c \in \topics}$. 
The algorithm terminates when $H_c^{\tau}$ converges.

Specifically, \myalgo iterates the two classification steps 
$\mbox{\sc UserClass}$  and $\mbox{\sc HashtagsClass}$.
%
%
\begin{algorithm}[t!]
\scriptsize
\caption{\label{alg:users-class} User Classification Algorithm}
\begin{algorithmic}[1]
\Require The set of polarized hashtags $H_c$ and the
previously found 
\Statex \hspace{2.5em}set of polarized users $U_c^{*}$ for each class $c \in \topics$
\Ensure New set of polarized users  $\{U_c\}_{c \in \topics}$
\Procedure{UsersClass}{\; $\{H_c\}_{c \in \topics}, \{ U_{c}^{*} \}_{c \in \topics} $\; }
\For{$u \in \users$, $c \in \topics$} \Comment{Find polarized tweets}
\State $T_{u,c} = \{t \in \tweets_u \mid \htags_t \cap H_c \neq \emptyset \wedge \htags_t \cap H_{c'\neq c} = \emptyset\}$\label{line:mod1}
\EndFor
\For{$c \in \topics$}
\State $U_c \gets \emptyset$
\EndFor
\For{$u \in \users$} \Comment{Check user's polarization}
\If{$\exists c \in \topics \mid \forall c'\in \topics,c'\neq c\quad |T_{u,c}|>\alpha\cdot|T_{u,c'}|$}\label{line:new}
\State $U_c \gets U_c \cup u$ 
\ElsIf{$ \exists c \in \topics \mid u \in U_{c}^{*} $}\label{line:backup}
\State $U_c \gets U_c \cup u$ 
\EndIf
\EndFor
\State {\bf return} $\{U_c\}_{c \in \topics}$
\EndProcedure
\end{algorithmic}
\end{algorithm}
%
%
Algorithm~\ref{alg:users-class} illustrates the former step of the iterative process\footnote{\scriptsize Note that we omitted the superscript $\tau$ for the sake of simplifying the notation.}.
The goal of this step is to identify polarized users on the basis of the given
hashtags. First, we identify polarized tweets, which mention hashtags in $H_c$. 
We consider the classification of each single tweet $t$ by considering all the mentioned 
hashtags $\htags_t$, as we believe each tweet is 
a very relevant expression of a user's thought on a specific topic. Since we
are interested in polarized users, with the goal of achieving high precision
we discard all the tweets which contain hashtags belonging to more than one set $\{H_c\}_{c \in \topics}$.
For each user $u \in \users$ and for each class $c \in \topics$ we denote the set of
polarized tweets by $T_{u,c}$. 
We thus measure the user polarization: if
for some classes $c$, the number of tweets in $T_{u,c}$ is significantly larger
than for any other class (parameter $\alpha$), then the user is labeled with the class $c$ and added
to the set of polarized users $U_c$ (see line \ref{line:new}). Note that the user classification
is intended to be an update of the classification conducted during the previous step.
%
%
%
%
%
%
The goal the second step is to process all the hashtags
adopted by classified users $U_c$ in order to discover a new set of
discriminating hashtags $H_c$, as illustrated in Alg.~\ref{alg:hashtags-class}. 
In order to detect $\{H_c\}_{c \in \topics}$, we take into considerations all the hashtags
$\htags_u$ used by any user $u \in U_c$, and not only those occurring in the
polarized tweets $T_{u,c}$ (line \ref{row:1}). This allows to extend our analysis to the full set of topics discussed by the users, even if they were not captured in the early
iterations of the algorithm.
\begin{algorithm}[t!]
\scriptsize
\caption{\label{alg:hashtags-class} Hashtag Classification Algorithm}
\begin{algorithmic}[1]
\Require The set of polarized users $U_c$ for each class $c \in \topics$
\Ensure Polarized hashtags $H_c$
\Procedure{HashtagsClass}{\; $\{U_c\}_{c \in \topics}$ \;}
\For{$c \in \topics$}
\State $H_c \gets \emptyset$
\State $H^*_c \gets  \bigcup_{u \in U_c} \htags_u$ \label{row:1}
\EndFor
\For{$h \in \bigcup_{c \in \topics} H^*_c$}
\If{$\exists c \mid \forall c'\neq c\quad S_c(h)> \beta \cdot S_{c'}(h) $} \label{line:1}
\State $H_c \gets H_c \cup h$ \label{line:2}
\EndIf
\EndFor
\State {\bf return} $\{H_c\}_{c \in \topics}$
\EndProcedure
\end{algorithmic}
\end{algorithm}
First, for each $c \in C$ we retrieve the set of hashtags used by the users in
$U_c$, considering all their tweets, denoted by $\tweets_c$, independent of the classification of the single tweets in the previous iteration. 
In our experiments we consider the top frequent 500 hashtags in $\tweets_c$.  
Given the resulting set of candidate hashtags for each $c \in \topics$, namely $H^*_c$, 
we extract from them the new hashtags that highly discriminate each class $c$, and these are eventually added to the new set $H_c$ 
(line \ref{line:2}). Specifically,
the discriminating hashtags are those highly used by the current set of users $U_c$, and partially used by any other user in $U_{c'}$, $c'\neq c$. 
We define a function $S_c(h)$ to measure the goodness of hashtag $h$ for each community of polarized users $U_c$.
Let
$\tweets_h$ be the set of tweets in $\tweets$ mentioning hashtag $h$, independent of the users who posted these tweets. Moreover, let
 $\tweets_{H^*_c}$ be the set of tweets in $\tweets$ containing at
least one hashtag in the set $H^{*}_c$. 
We score the goodness of a hashtag for a polarization class as follows:
\begin{equation} 
\nonumber\textstyle 
S_c(h) = \frac{ |\tweets_h \cap \tweets_{H^*_c}| }{ |\tweets_{H^*_c}| } \cdot \prod_{c' \in \topics, c'\neq c} \left( 1 - \frac{ |\tweets_h \cap \tweets_{H^*_{c'}}| }{ |\tweets_{H^*_{c'}}| } \right)
\end{equation}
\noindent where we consider the naive hypothesis of independent occurrence of the hashtags in the various sets.
In practice, $S_c(h)$ is the probability of seeing $h$ only in $H^{*}_c$, whereas $h$ is not present in all the other sets of hashtags $H^{*}_{c' \neq c}$.
Given a hashtag $h$, the score $S_c(h)$ is used to rank the various classes, 
thus assigning $h$ to class with the highest score.
Since we aim at promoting high discriminating hashtags, not only we assign  
the hashtag $h$ having the highest $S_c(h)$ to the new set $H_c$, but 
only if $S_c(h) > \beta \cdot S_{c'}(h)$, $\forall c'\neq c$, where $\beta \ge 1$.
Note that if a tie exists between the to 2-top scores classes, the hashtag $h$ is not assigned to any $H_c$, since it is considered not discriminating enough.


\section{Experimental Evaluation}
\label{sec:experiments}

\subsection{Data collection and cleansing}


We use two Twitter datasets related to
political elections that recently took place in Italy.
\textbf{Dataset \pd}: data about primary election for largest
social democratic political party in Italy (PD), which took place in December 2013 with 3 candidates: Mr.\
Renzi, Mr.\ Cuperlo, and Mr.\ Civati. 
%
\textbf{Dataset \eu}: data about European Parliament election held in Italy in May 2014\footnote{\scriptsize The main national parties connected to
different European political groups were: \textit{Partito Democratico} (PD),
\textit{Movimento 5 Stelle} (M5S), \textit{Forza Italia} (FI), \textit{Lega Nord} (LN), \textit{Tsipras} (AET). We ignore smaller parties and NCD-UDC for its limited presence in Twitter.
}.
%
\begin{table}
\caption{Data Statistics}
\centering
\scriptsize
\subfloat[Full dataset]{
\label{tab:datainfo} 
\begin{tabular}{lcc}
\textbf{Dataset} & \textbf{\datapd} & \textbf{\dataeu} \\ 
\hline\hline
tweets in original raw data & 1.7 M. & 2.3 M.\\
pre-electoral tweets \tweets & 95,627 & 364,132\\
users with $|\htags_u|>0$ & 11,368 (65\%) & 28,340 (56\%)\\
\hline
\end{tabular}
}
\quad
\subfloat[Golden dataset]{
\label{tab:datainfoa}
\begin{tabular}{lrr|lrr}
\multicolumn{3}{c|}{\bf Dataset \datapd} &  \multicolumn{3}{c}{\bf Dataset \dataeu} \\
$\topics$ & Tweets & Users & $\topics$ & Tweets &  Users\\
\hline
\hline
Renzi 		& 330 & 109	& PD 		& 262 & 129 	\\
Cuperlo 	& 4759 & 243 		& M5S 		& 146 & 95 	 \\
Civati 		& 2925 & 700 	& FI 		& 1263 & 199 	 \\
	 		 & 	 	& 			& LN 		& 480 & 226  \\
			 &		&			& AET 	& 757 & 328 	\\ \hline
{\it total}	& 8014 & 1052				& {\it total}	& 2908 & 977  \\
\end{tabular}
}
\end{table}
%
%
%
The data are collected through Twitter API by querying a list of keywords
related to the topic and the candidates, large enough to
guarantee a good coverage of the elections. Both final datasets cover 9 days before the election day.  
We discard partial data and potentially
irrelevant tweets, considering only tweets being geo-located and in Italian language. 
Table~\ref{tab:datainfoa} reports some information about the two datasets.

\subsection{Evaluation of the \myalgo algorithm}

We build an evaluation dataset by identifying those users whose opinion can be
inferred with high confidence. During elections, as for other events, very
specific hashtags are used over Twitter to express a strong intention of vote
or an explicit membership in a group. We assume that users that frequently use
one of such hashtags are strongly sided with one of the competing parties and
they will not change idea in the short term. Such hashtags, named {\em golden
hashtags}, are handpicked among the 500 most frequent in the data. 
The used golden hashtags are of the kind
{\tt \#IVoteParty}. We identify one/two golden hashtags per class $c \in \topics$ both in the \dataeu (e.g. {\tt \#IVoteTsipras} for AET) and in the \datapd (e.g. {\tt \#prefeRenzi} for Renzi) dataset.
The set of reference users were identified by applying Algorithm~\ref{alg:users-class} 
with the above {\em golden} hashtags as input. This guarantees that a user is
safely considered as polarized to a party $c \in \topics$ if her tweets 
contain only one of the golden hashtags associated with 
the various classes $c \in \topics$.
We denote with $\cusers=\{z_1, z_2, \ldots \}$ this set of polarized users,
and with $Z_c\subseteq \cusers$ those supporting a specific formation $c$ 
($Z_c$ is a partitioning of \cusers).
The composition of resulting {\em golden dataset} is reported in
Table~\ref{tab:datainfo}. 
The golden dataset is thus a small fraction of
the full dataset.
A global analysis of the
Twitter stream cannot be based on a few very polarized hashtags. Note that the
relative popularity of the parties is not simply proportional to the number of votes
received, but it depends on the efficacy of the hashtag promoted.
We remark that, for the sake of fairness, we remove the {\em golden} hashtags from
the datasets before the application of any algorithm. 
%
The set of users \cusers in the golden dataset, is used to evaluate
the users classification accuracy of the proposed method. Given the users
classification $U_c$ provided by some given algorithm, precision, recall and
F-Measure are restricted to the set \cusers. Formally, for any given class
$c \in \topics$, precision and recall are defined as:
\begin{equation}
\nonumber\textstyle 
P_c (U_c) =  \frac{ |U_c \cap Z_c|}{|U_c \cap \cusers|} \quad\quad
R_c (U_c) =  \frac{ |U_c \cap Z_c|}{|Z_c|}
\end{equation}
The F-measure $F_c$ is the harmonic means of $P_c$ and $R_c$. The macro $F$-measure average over the
classes $c \in \topics$ is denoted with $F$.
In addition, as the proposed algorithm may not be able to classify all of the users in \cusers,
we report also the user coverage $\gamma$ and $\Gamma$ on both the golden set and the overall dataset respectively:
\begin{equation}
\nonumber\textstyle 
\gamma(U=\cup_{c \in \topics} U_c )= \frac{ |U \cap \cusers| }{ |\cusers| } \quad\quad
\Gamma(U=\cup_{c \in \topics} U_c )= \frac{ |U| }{ |\users| } 
\end{equation}

As a baseline we use the $k$-means clustering algorithm. Each user $u$ is represented by a vector of 500 features,
corresponding to the 500 most frequent hashtags in the dataset. The user
feature vector stores the frequency of a hashtag in the stream of tweets
$\tweets_u$ published by the user. We discard users who do not use any
hashtag in their tweets.
We normalize the feature vectors for each
user to unit $L^2$ norm. We impose the number of the clusters $k$ equal to
the number of classes $|\topics|$ and, to simulate the same starting condition of our
method, we built the initial centroids so as to encode the {\em seed} hashtags.
The centroid for a class $c$ is thus a vector
with a single 1 in the position of the seed hashtag, and 0 otherwise. The result of
the $k$-means baseline is thus a clustering of users based on the {\em seed}
hashtags provided.
Table~\ref{tab:kmeans} reports the results of the $k$-means baseline.
F-measure values are low for the \datapd dataset. For instance, $k$-means
provides low accuracy and recall for the first class. This is mainly due to
the fact that the hashtags corresponding to popular parties or candidates are
very often used by different users, regardless of their orientation. In other
cases (e.g., LN and AET), the hashtags are
used mostly within the respective communities. 
%
\begin{table}[b!]
\caption{Comparison with the Baseline.}
\centering
\scriptsize
\subfloat[$k$-means baseline performance]{
\label{tab:kmeans}
\STautoround*{3} 
\begin{spreadtab}{{tabular}{@{}l@{\ \ }rrr|l@{\ \ }rrr}}
\multicolumn{4}{c|}{@ {\bf Dataset \datapd}} &  \multicolumn{4}{c}{@ {\bf Dataset \dataeu}} \\
@\topics & @$P_c$ & @$R_c$ & @$F_c$ & @\topics & @$P_c$ & @$R_c$ & @$F_c$ \\
\hline
\hline
@Renzi & 0.144 & 0.257 & 0.185 		& @PD 	& 0.536 & 0.457 & 0.493	\\
@Cuperlo & 0.252 & 0.543 & 0.344	& @M5S 	& 0.359 & 0.895 & 0.512 \\
@Civati & 0.766 & 0.366 & 0.495		& @FI 	& 0.495 & 0.734 & 0.591 \\
	&	&	&						& @LN 	& 0.995 & 0.916 & 0.954 \\
	&	&	&						& @AET 	& 1.000 & 0.387 & 0.558 \\ \hline
@{\it avg.} &  sum(b3:b5)/3	& sum(c3:c5)/3 & sum(d3:d5)/3 &
@{\it avg.} &  sum(f3:f7)/5	& sum(g3:g7)/5 & sum(h3:h7)/5 \\ \hline \hline
\multicolumn{2}{r}{@ $\gamma=1.0$} & \multicolumn{2}{r|}{@ $\Gamma=0.653$} & 
\multicolumn{2}{r}{@ $\gamma=1.0$} & \multicolumn{2}{r}{@ $\Gamma=0.557$}
\end{spreadtab}
}\quad
\subfloat[\myalgo Iteration-2 performance]{
\label{tab:step2}
\STautoround*{3} 
\begin{spreadtab}{{tabular}{l@{\ \ }rrr|l@{\ \ }rrr@{}}}
\multicolumn{4}{c|}{@ {\bf Dataset \datapd}} &  \multicolumn{4}{c}{@ {\bf Dataset \dataeu}} \\
@\topics & @$P_c$ & @$R_c$ & @$F_c$ & @\topics & @$P_c$ & @$R_c$ & @$F_c$ \\
\hline
\hline
@Renzi 		& 0.350 & 0.752 & 0.478 	& @PD 	& 0.733 & 0.488 & 0.586	\\
@Cuperlo 	& 0.869 & 0.300 & 0.446		& @M5S 	& 0.325 & 0.842 & 0.469 \\
@Civati 	& 0.916 & 0.747 & 0.823		& @FI 	& 0.955 & 0.533 & 0.684 \\
	&	&	&							& @LN 	& 0.981 & 0.938 & 0.959 \\
	&	&	&							& @AET 	& 0.974 & 0.451 & 0.617 \\ \hline
@{\it avg.} &  sum(b3:b5)/3	& sum(c3:c5)/3 & sum(d3:d5)/3 &
@{\it avg.} &  sum(f3:f7)/5	& sum(g3:g7)/5 & sum(h3:h7)/5 \\ \hline \hline
\multicolumn{2}{r}{@ $\gamma=0.845$} & \multicolumn{2}{r|}{@ $\Gamma=$ 0.532} & 
\multicolumn{2}{r}{@ $\gamma=0.830$} & \multicolumn{2}{r}{@ $\Gamma=$ 0.367}
\end{spreadtab}
}
\end{table}

In the following, we analyze in detail the iteration-by-iteration behavior of
the proposed \myalgo algorithm. We test our algorithm by setting $\alpha=2$
and $\beta=1$, after a tuning step.
During the first iteration, \myalgo is fed with the {\em seed} hashtags.  Algorithm~\ref{alg:users-class} uses
those hashtags to find a subset of polarized users in \users. This step is
similar to other works, where mentions of a party or candidate are used to
estimate their popularity or to classify users
\cite{coletto,tumasjan2010predicting}. Unlike other approaches, 
\myalgo aims at discovering a subset of polarized users, thus requiring,that
a user mentions a party at least twice any other. The results of such
user classification are evaluated over the {\em golden dataset}, as reported
in the first line of Table~\ref{tab:iterations}. Regarding average precision,  \myalgo 
is already significantly superior to the $k$-means baseline for \datapd dataset.
This is already surprising, as the seed hashtags are very generic. On the other
hand, the $k$-means baseline might be negatively affected by the sparsity of
the data. The results are different on the two datasets in terms of average
recall. \myalgo has similar performance to $k$-means on the \datapd dataset,
while the recall is significantly lower on the \dataeu dataset. This is
confirmed  by the coverage values $\gamma$ and $\Gamma$.
In comparison with the baseline, the performance of \myalgo in terms of macro
$F$-measure is satisfactory on the \datapd dataset, but not on the
\dataeu dataset yet.
The output of the first iteration is a new set of hashtags which is exploited
in the next iteration. By looking at the best scoring hashtag, we can already
observe an interesting behavior of the algorithm for some $c \in \topics$. In dataset \dataeu, the best tags for FI and LN are the leaders of the respective parties, detecting that the original {\em seed} hashtags are not discriminating in this case. 
In Table~\ref{tab:step2} we report in detail the results after the second iteration
of \myalgo. The first interesting result is that the average recall is
significantly higher on both datasets. This is due to the new hashtags
discovered in addition to the {\em seed} ones during the previous iteration,
which, in turn, lead to the identification of a larger set of users: the coverage
$\gamma$ is now beyond 80\% on the {\em golden} set, and $\Gamma$ has doubled
in this iteration. 
Also the
average precision is higher w.r.t.\ the previous iteration scoring more than
0.7. This is both because of the increased number of classified users, and of
the updated user classification. 
As a result, the $F$-measure has an overall improvement w.r.t.\ the $k$-means
baseline of +71\% and +7\% on datasets \datapd and \dataeu respectively.
\begin{table}
\caption{Algorithm Performance.}
\centering
\scriptsize
\subfloat[\myalgo iteration by iteration performance]{
\label{tab:iterations} 
\begin{tabular}{c|rrr|rrr}
 & \multicolumn{3}{c|}{\bf Dataset \datapd} &  \multicolumn{3}{c}{\bf Dataset \dataeu} \\
Iter & $F$ & $\gamma$ & $\Gamma$ & $F$ & $\gamma$ & $\Gamma$ \\
\hline
\hline
1  	& 0.358 & 0.490 	& 0.218 & 0.514 & 0.670 & 0.163	\\
2	& 0.582 & 0.845 	& 0.522 & 0.663 & 0.830 & 0.367	\\
3 	& 0.588 & 0.853 	& 0.532 & 0.662 & 0.831 & 0.386 \\
4	& 0.588	& 0.853		& 0.534 & 0.661 & 0.834 & 0.390 \\ \hline
\end{tabular}}
\quad
\subfloat[\myalgotime day-by-day performance]{
\label{tab:dailyiterations} 
\vspace{\baselineskip}
\begin{tabular}{c|ccc|ccc}
 & \multicolumn{3}{c|}{\bf Dataset \datapd} &  \multicolumn{3}{c}{\bf Dataset \dataeu} \\
Day & $F$ & $\gamma$ & $\Gamma$ & $F$ & $\gamma$ & $\Gamma$ \\
\hline
\hline
1  	& 0.177 & 0.199 	& 0.045 & 0.155 & 0.164 & 0.025 	\\
2	& 0.225 & 0.348 	& 0.114 & 0.464 & 0.465 & 0.079	\\
3 	& 0.304 & 0.457 	& 0.166  & 0.529 & 0.570 & 0.116 \\
4	& 0.333	& 0.563	& 0.234 & 0.585 & 0.671 & 0.180 \\ 
5  	& 0.368 & 0.606 	& 0.261 & 0.588 & 0.726 & 0.235	\\
6	& 0.397 & 0.671 	& 0.315 & 0.574 & 0.762 & 0.269	\\
7 	& 0.387 & 0.721 	& 0.363 & 0.596 & 0.794 & 0.302 \\
8	& 0.387	& 0.765	& 0.408 & 0.637 & 0.846 & 0.334\\
9	& 0.391	& 0.811	& 0.461 & 0.635 & 0.876 & 0.349 \\\hline
\end{tabular}
}
\end{table}
As shown in Table~\ref{tab:iterations} 
\myalgo 
becomes stable very early. 
The largest improvement is achieved with the second
iterations. This means that the most relevant hashtags are discovered early,
and only slight changes occur afterwards. The subsequent iterations marginally
increase the number of classified users. Note that the algorithm is
classifying the polarized users found in the whole set \users. \myalgo found
about  6.7 and 27 thousands polarized users on the dataset \datapd and \dataeu
respectively. We conclude that in most cases, two iterations of the algorithm
provide sufficient classification quality. For the lack of space we can not report a exhaustive qualitative analysis of the outcome, but we observe that the procedure is able to extract relevant keywords: namely prominent politicians, the party itself and political mottoes characterizing each $c$ in the political scene.

We finally propose a variant of \myalgo, that is \myalgotime (temporal \myalgo), to perform the tracking of topics and users in time. In our case we consider the evolution day by day.
The procedure follows Algorithm~\ref{alg:users-class} and Algorithm~\ref{alg:hashtags-class} with the difference that at iteration 
$\tau$ only the tweets $\tweets_u$ written in the $\tau$-th day are considered.
%
%
%
We perform \myalgotime on \pd and on \eu datasets.
In Table~\ref{tab:dailyiterations} the evaluation of the temporal iterative procedure is shown.
The macro $F$-measure is increasing day by day both for the effect of a better classification and for the presence of new users. 
Note that we evaluate the time iterative method day by day on the entire \textit{golden set} of users.
F-measure values are low because not all users in the \textit{golden set} were active every day. 

\section{Conclusion}
\label{sec:conclusion}

We propose a novel algorithm for the simultaneous tracking of \emph{polarized users} and \emph{discriminating topics} in OSNs. Specifically, it iteratively detects polarized users, and from their contents the discussed discriminating topics. 
We also introduce a
temporal variant, where the information extracted during one day of analysis
is exploited for the next day. Indeed, the classification of users makes
the algorithm more robust in terms of concept drifts, as new trends may be
detected as early as they pop up. At the same time, the
identification of discriminating topics helps in 
detecting users moving from one
class to another.
The algorithm is tested on two Twitter data samples. We evaluate the quality of user classification on a 
\textit{golden set} of users, showing significant improvements
over the baseline.
The proposed methodology is general and it can be applied to different scenarios. 
We
believe that this methodology based on {\em polarization} may also impact
on broad area of social network analysis, e.g., by complementing the proposed
classification with community detection and
information diffusion over time.
As a future work, we aim to improve the temporal analysis dealing with streaming data.

\noindent {\footnotesize \textit{Supported by EC H2020 Program INFRAIA-1-2014-2015 (654024).}}

\bibliographystyle{abbrv}
\bibliography{biblio}
%


\balancecolumns

\end{document}